\documentstyle[12pt,aaspp4,flushrt]{article}
\journalid{337}{15 January 1989}
\articleid{11}{14}

\def\etal{{et al.}}
\begin{document}

\title{Stars of the Draco Dwarf Spheroidal Galaxy Beyond its Measured
Tidal Boundary}
\author{Slawomir Piatek}
\affil{Dept. of Physics, New Jersey Institute of Technology, Newark,
NJ 07102 \\
     E-mail address: piatek@physics.rutgers.edu}

\author{Carlton Pryor\altaffilmark{1}}
\affil{Dept. of Physics and Astronomy, Rutgers, the State University of
New Jersey, 136~Frelinghuysen Rd., Piscataway, NJ 08854--8019 \\
E-mail address: pryor@physics.rutgers.edu}

\author{Taft E.\ Armandroff}
\affil{National Optical Astronomy Observatory, P.O.~Box~26732, Tucson, AZ 85726 \\
     E-mail address: tarmandroff@noao.edu}
\and

\author{Edward W.\ Olszewski\altaffilmark{1}}
\affil{Steward Observatory, University of Arizona,
    Tucson, AZ 85721 \\ Email address: eolszewski@as.arizona.edu}

\altaffiltext{1}{Visiting Astronomer, Kitt Peak National Observatory,
National Optical Astronomy Observatory, which is operated by the
Association of Universities for Research in Astronomy, Inc. (AURA)
under cooperative agreement with the National Science Foundation.}

\begin{abstract}
We report $R$-- and $V$--band photometry derived from CCD imaging for
objects in nine fields in and around the Draco dwarf spheroidal
galaxy.  The most distant fields are about 1.3$^\circ$ from the
center.  We use these data to search for Draco stars outside of its
measured tidal boundary.  The search involves three methods:  1)
Plotting color-magnitude diagrams for individual fields, for sections
of fields, and for combined fields and sections.  A color-magnitude
diagram can reveal a population of Draco stars by the presence of
the expected principal sequences.  2) Measuring field-to-field
fluctuations and 3) measuring intra-field fluctuations in the
surface density of objects located near the Draco principal sequences
in the color-magnitude diagram.  We find evidence for the presence of
Draco stars immediately beyond the measured tidal boundary of Draco and
place an upper limit on the number of such stars in more distant fields
that lie close to the extension of its major axis.  The best evidence
is the presence of the Draco principal sequences in the color-magnitude
diagram for some combined fields and sections of fields.  The
measurements of the field-to-field fluctuations in the stellar surface
density confirm this result.
\end{abstract}

\section{Introduction} \label{intro}

	It is a long-standing question whether the tidal forces of the
Milky Way have played, and continue to play, a significant role in the
formation and evolution of its companion dwarf spheroidal (dSph,
hereafter) galaxies (e.g., Hodge \& Michie 1969).  Proposed effects of
these forces on a dSph are episodic star formation (Lin \& Murray
1998), a spuriously large measured mass-to-light ratio (e.g., Kuhn \&
Miller 1989; Bellazzini \etal\ 1996; Klessen \& Kroupa 1998; but also
see Pryor 1996; Olszewski 1998; Hirashita \etal\ 1999), and structural
alterations of the outer regions.  Examples of structural alterations
are isophote twisting, truncation of the radial density profile, and
the formation of extended streams of tidal debris.  In this article we
look for the last of these effects using star counts in two colors to
search for stars of the Draco dSph beyond its measured tidal boundary.

Irwin \& Hatzidimitriou (1995; hereafter IH) list the most accurately
measured value of the ``tidal radius'' of Draco: $28.3 \pm 2.4$~arcmin
along the major axis.  They obtained this value by fitting a
single-component King (1966) model to their projected density profile
derived from star counts made using Palomar Schmidt plates digitized and
analyzed using the APM facility (Kibblewhite \etal\ 1984).
Because there is no fundamental reason why a King model must describe a
dSph and, as IH note, the models do not fit the outer parts of their
profiles well, this measured tidal radius should not be literally
interpreted either as the limiting radius of a dSph or as the radius
beyond which stars are unbound due to Galactic tides.  Thus, this tidal
radius should only be considered as a parameter in a fitted model.
Throughout this article we use the phrase ``tidal boundary'' to mean
the elliptical contour centered on Draco with a semi-major axis of
28.3~arcmin, an ellipticity of 0.29, and position angle of 82~degrees
(IH).  However, we do not automatically interpret Draco stars beyond
this boundary as being unbound from Draco by Galactic tidal forces.

Even though the physical interpretation of the tidal radius resulting
from fitting a King model to the projected density profile of a dSph is
uncertain, the spatial structure of the outer regions of a dSph is an
important source of information about the galaxy.  The density profile
of bound stars at large radii is needed to constrain the central
density of dark matter in a dSph (Pryor 1994).  Moore (1996) has
emphasized that the radius beyond which most of the stars are unbound
due to Galactic tidal forces constrains the total mass of a dSph and,
hence, its total amount of dark matter.  Johnston \etal\ (1999)
identify this radius with that where the slope of the projected density
profile in a log-log plot abruptly becomes shallower, based on
numerical simulations of dSphs interacting with the tidal field of the
Galaxy.  The unbound stars also should show velocity gradients and a
velocity dispersion increasing with increasing distance from the dSph
(Oh \etal\ 1995; Piatek \& Pryor 1995; Johnston \etal\ 1999).
Observing these kinematic signatures is necessary to confirm that these
stars are unbound from the dSph by Galactic tidal forces, since other
mechanisms could produce an extended bound population with a shallow
density profile at large radii.  An example is a halo of stars that
formed from gas pushed to large radii by supernovae and stellar winds.

	In this article, we report $R$-- and $V$--band photometry for
objects in nine fields in and around Draco and the results of using
these data to search for Draco stars beyond its measured
tidal boundary.  Section 2 describes the data acquisition and
reduction.  Section 3 presents color-magnitude diagrams for our fields
and discusses the classification of an object based on its location in
the diagram and the morphology of its image.  Section~4 describes our
methods of searching for Draco stars and the results they yield.
Section 5 summarizes our results and compares them to the results from
other similar studies.

\section{Data Acquisition and Reduction} 
\label{data}

	We imaged nine fields in and around Draco in the R and V bands
using the KPNO 0.9-m telescope and two $2048\times2048$ pixel CCD
chips: an engineering grade chip for the central field -- reading out
only 1747 columns -- and the CCD designated T2KA for the rest.  The
image scale is 0.68~arcsec per pixel, yielding 23.2~arcmin square
images for all but the central field.  Table~1 contains the basic
information about the fields. Column 1 lists their abbreviated
names. Column 2 gives the date when each was imaged. Columns 3, 4, 5,
and 6 list the equatorial and Galactic coordinates of the centers of
the fields, respectively. Columns 7 and 8 list the combined exposure
times in the $V$ and $R$ bands and the final two columns list the
average full-width at half-maximum (FWHM), in pixels, measured for
stellar images in the $V$ and $R$ frames.  This value of the FWHM is
an average of those for nine stars forming a rough $3\times3$ grid
across a field and is followed by the root-mean-square (rms,
hereafter) scatter around the average.  Generally, the seeing in the
$R$ band is better than that in the $V$ band, that for the C0 and W2 fields
being the exceptions.  The scatter, indicative of intra-field focus
variations, is generally small, except for the C0 and E1 fields.
The smaller scatter for the other fields is due to a two-element
corrector, installed between our 1993 and 1994 runs, which flattened
the focal surface of the 0.9~m telescope (see Armandroff \etal\ 1993).
Only in E1 are these variations large enough to cause problems with the
photometry, which are discussed later in this section.

Figure~1 shows the configuration of the fields superimposed on a
Digitized Sky Survey image centered on Draco.  The ellipse depicts the
measured tidal boundary of Draco; its semi-major axis and ellipticity
are 28.3~arcmin and $0.29$, respectively (IH).  The tidal boundary
encloses all of the C0 field, more than $50\%$ of the E1 and W1 fields,
and only about $10\%$ of the SE1 field -- the northwest corner.
Figure~2 shows the fields superimposed on a map of the surface density
of stars and galaxies derived by IH for a $2.3^\circ \times 2.3^\circ$
region centered on Draco.  Figures~1 and 2 imply that fields C0, E1,
W1, and SE1 contain Draco stars.

The data for each Draco field generally consists of four exposures in
the $V$ band and three in the $R$ band (see columns 7 and 8 of Table~1
for details).  We overscan subtracted, flattened, and trimmed each
image in the usual way using IRAF\footnote{IRAF is distributed by the
National Optical Astronomy Observatory, which is operated by the
Association of Universities for Research in Astronomy, Inc., under
contract to the National Science Foundation.}.  For the 1994 data,
twilight flats produced the most uniform sky for the $R$ images whereas
dome flats produced the most uniform sky for the $V$ images.  For
similar reasons, we used twilight flats for the 1993 (E1 field) $V$
images and dome flats with an illumination correction based on the
twilight sky for the $R$ images.  The 1991 C0 field had only dome flats
available.  Our next processing step was aligning and then combining
the frames in a given band and field to form a single image.  Our
primary goal in combining frames was to eliminate contamination due to
cosmic rays, satellites, meteors, etc..  The optimal rejection of
deviant pixels requires consistent sky levels and stellar fluxes in the
different frames. We multiplied each frame by the amount necessary to
make the aperture fluxes of three bright, isolated stars equal, on
average, to their fluxes averaged over all of the images. Additive
corrections then made the average skies measured in the vicinity of
these three stars equal. We averaged the frames using the IRAF
IMCOMBINE subroutine using the ``ccdclip'' rejection algorithm with a
$4\sigma$ rejection limit.

	We used the stand-alone version of the photometry software
package DAOPHOT (Stetson 1987, 1992, 1994) to find objects in the
combined frames and to determine their magnitudes by fitting a point
spread function (PSF, hereafter) derived from selected stars.
About 100 stars determined the PSF on each averaged frame and the PSF varied
quadratically with position within a frame. The DAOPHOT ALLSTAR
program fits the PSF to each object found in the averaged frame, yielding
the centroid, magnitude, uncertainty in the magnitude, sky
level, CHI, and SHARP. The last two parameters measure how closely the
fitted image matches the PSF. CHI is a robust measure of $\chi^2$
and SHARP describes whether the image is more or less extended than
the PSF.  If the SHARP is close to 0, the
object is likely to be a star; if it is significantly less than 0, the
object is probably a corrupted pixel; and, if it is significantly larger
than zero, the object is likely to be a galaxy or unresolved double
star.

	We matched the objects photometered in the $R$ and $V$ frames
and determined the coordinate transformation between the frames.  We
derived the instrumental magnitudes with ALLFRAME (Stetson 1994), which
simultaneously fits a PSF to objects in both colors using single
positions and the coordinate transformation between frames.  The input
to ALLFRAME is a list of objects found in at least one frame.  If
ALLFRAME does not detect an object in both frames simultaneously, we
eliminate it from further consideration.

        The next step in the image reduction procedure is
converting the instrumental magnitudes, $R_{\rm instr}$ and $V_{\rm
instr}$, to standard $R$ and $V$ magnitudes.  We define aperture
magnitudes, $R_{\rm ap}$ and $V_{\rm ap}$, which are measured using the
IRAF PHOT command with an aperture radius of 20 pixels and a sky
annulus with inner and outer radii of 20 and 35 pixels.  Curves of
growth showed that this aperture is sufficiently large to include a
constant fraction of the light independent of the known variability of
the PSF.  The instrumental, aperture, and standard magnitudes are
related through the following equations:
\begin{equation}
V_{ap}=V_{instr}-a^{\prime}=V+a+b(V-R)+cX
\label{convertV}
\end{equation}
and
\begin{equation}
R_{ap}=R_{instr}-d^{\prime}=R+d+e(V-R)+fX^{\prime},
\label{convertR}
\end{equation}
where $X$ and $X^{\prime}$ are the average airmasses for the combined
frames. The constants $a^{\prime}$ and $d^{\prime}$ are the zero-point
offsets between the instrumental and aperture magnitudes for the $V$
and $R$ bands, respectively.  Similarly, $a$ and $d$ are zero-point
offsets between the aperture and standard magnitudes, $b$
and $e$ are the color correction coefficients, and $c$ and $f$ are the
extinction coefficients.

	To determine $a^{\prime}$ and $d^{\prime}$ for a given field,
we measured aperture magnitudes with a radius of 20 pixels for the PSF
stars on frames in which the neighbors of the PSF stars were subtracted
(thus minimizing contamination).  The average difference between the
aperture magnitudes of the PSF stars and their corresponding
instrumental magnitudes yields the values of the two offsets.

	On each night of the photometric 1994 observing run, we imaged
33 standard stars (Landolt 1983, 1992) with $9.3 < V < 16.1$ and $0.1 <
V-R < 1.5$.  A least-squares fit of equations~\ref{convertV} and
\ref{convertR} to the aperture magnitudes of the standards determined
the coefficients of the transformation between aperture and standard
magnitudes.  Table~2 lists the values of the coefficients $a$ through
$f$ which produce the best fit for a given night.  We adopted the
average color correction and extinction coefficients for the run, but
use separate zero-points for each night.  Adopting average zero-points
resulted in unacceptably large scatter around the fitted relations.
Substituting these coefficients and the average airmass values for a
given field and band from Table~3 into equations~\ref{convertV} and
\ref{convertR} yields the standard magnitudes for the detected objects
in all but the C0 and E1 fields, which were imaged during the
non-photometric 1991 and 1993 observing runs, respectively.

	On the second night of the 1994 observing run, we imaged in
both $V$ and $R$ a tie field, E1(tie) (see Table~1), which overlaps the
C0 and E1 fields. We identified the C0 and E1 PSF stars contained in
the E1(tie) field, measured their aperture magnitudes in the E1(tie)
frame, and derived their standard magnitudes with
equations~\ref{convertV} and \ref{convertR}, setting the fitting
coefficients $a$ through $f$ to the night 2 values (see Table~2).  The
instrumental magnitudes measured in the C0 and E1 fields for these
selected PSF stars are also related to their derived standard
magnitudes through equations~\ref{convertV} and \ref{convertR}
(combining $a$ and $a^\prime$ and $d$ and $d^\prime$ each into a single
constant).  The least-squares fit of equations~\ref{convertV} and
\ref{convertR} to these data yields sets of coefficients appropriate
for the C0 and E1 fields.  Substituting these coefficients in
equations~\ref{convertV} and \ref{convertR} yields standard magnitudes
for the detected objects in the C0 and E1 fields.

	Tables~4(a) through 4(i) present our final $R$- and $V$-band
photometry for the C0, E1, SE1, W1, N2, E2, S2, SW2, and W2 fields,
respectively.  Only the beginning of Table~4(a) is shown in the
printed version of this article to provide guidance on form and
content.  The entire tables appear in the electronic edition of the
\textit{Astronomical Journal}.  The first five columns in the tables
give the ID, $x$ and $y$ coordinates on the $R$ frame, and $\alpha
({\rm J2000.0})$ and $\delta ({\rm J2000.0})$ for an object.  The equatorial
coordinates come from plate solutions based on positions for stars in
the USNO-A2.0 catalog (Monet \etal\ 1998) using a recipe developed
by Paul Harding (personal communication).  Columns 6 and 7 give the
$R$-band magnitude followed by its uncertainty, $\sigma_{R}$. Columns 8
and 9 list the same for the $V$ band. The last two columns give the
average CHI and SHARP values.

\subsection{Comparison to Earlier Photometric Studies of Draco}
\label{photcomp}

	There are two ground-based CCD photometric studies of Draco in
the literature: Stetson \etal\ (1985; hereafter SVM)
and Carney \& Seitzer (1986; hereafter CS).  Both groups imaged fields
completely contained within our C0 field and both observed in the $B$
and $V$ bands. Therefore, we can only compare the $V$-band photometry
from these studies to our photometry.

	SVM found 182 objects in their Draco fields; we matched 95 of
these objects within a limit of 1 pixel. Figure~3 plots the magnitude
difference $\Delta V \equiv V_{\rm us}-V_{\rm SVM}$ vs. $V_{\rm SVM}$,
where $V_{\rm us}$ is the magnitude determined in this study and
$V_{\rm SVM}$ is the magnitude of the same object determined by SVM.
CS imaged two fields in Draco: Field~1, where they found 1039 objects,
and Field~2, where they found 875.  We found 755 and 646 objects in
common between our C0 field and Fields 1 and 2, respectively, within a
limit of 1 pixel. We combined the two sets of objects into a single set
from which we excluded those with $V<18.0$ because CS note that their
photometry of these objects is unreliable due to saturation.  The final
set consists of 1382 objects; however, since CS Fields~1 and 2
overlap by 26\%, some objects appear twice in the final set.
Figure~4 plots the magnitude difference, $\Delta V \equiv
V_{\rm us} - V_{\rm CS}$ vs. $V_{\rm CS}$, for 1382 objects. $V_{\rm
CS}$ corresponds to the $V$--band magnitude determined by CS.

	Figures~3 and 4 show small systematic offsets between the
zero-points of the three studies for bright objects and a bias that
causes $\Delta V$ to become progressively more negative with increasing
magnitude starting at $V \simeq 21.0$.  The unweighted mean
offsets for objects brighter than $V = 21.0$ in Figures~3 and 4 are
$\Delta V = 0.009 \pm 0.019$ and $\Delta V = -0.008 \pm 0.004$,
respectively.  The uncertainty in the mean offset is the rms scatter
around the mean divided by the square root of the number of objects.
For the data in Figure~4, the calculation of the mean offset and rms
excluded the few points marked with a slanted cross.  These correspond
to some confirmed RR Lyrae stars from Baade \& Swope (1961).  However,
we have made no attempt to identify all of the known RR~Lyrae stars in our
fields.  The rms scatter around the mean offsets for the bright objects
is 0.068~mag for the comparison with SVM and 0.061~mag for that with
CS.  The corresponding values of the $\chi^2$ per degree of freedom are
$4.472$ and $2.375$.  These are larger than one at least in part
because of unidentified variable stars, close pairs of stars resolved
in one study and not the other, and galaxies photometered as stars.

The magnitude-dependent bias for fainter objects causes our magnitudes
to be smaller on average than those determined by SVM and CS for
corresponding objects.  The direction of the bias is the same in both
Figures~3 and 4, implying that its cause lies with our faint $V$-band
C0 data.  The SVM and CS data go much deeper than ours because they used 4~m
telescopes.  Near the faint limit of our photometry, we preferentially
detect those objects made brighter by noise, but not those made
fainter.  This process produces a bias in the direction of that seen in
Figures~3 and 4.  While this process probably produces most or all of
the bias for the faintest objects, it probably does not produce the
bias for objects brighter than about magnitude 22.  In this magnitude
range, examining some cases suggests that close pairs of stars
unresolved in our study cause the bias.  We note that the C0 field has
the shallowest photometry of any of our fields (see Sec.~\ref{complete}
below).

\subsection{Comparison of Photometry from Overlapping Fields}
\label{ftfcomp}

	Three pairs of our fields slightly overlap, E1~and~C0, E1~and~SE1,
and S2~and~SW2, allowing a comparison of photometry in both the $R$ and
$V$ bands.

	Figure~5(a) plots the magnitude difference $\Delta R \equiv
R_{\rm E1} - R_{\rm C0}$ vs.\ $R_{\rm E1}$ for the 190 objects common
to the E1 and C0 fields, matched to within 1 pixel.  The figure shows
a zero-point offset that extends over the entire range of magnitudes.
Field E1 has the most variable focus among our images, as noted when
discussing the stellar FWHMs in Table~1.  Most of the objects showing
the largest magnitude differences are located on the side of the
overlap with the worst focus.  At least one other large difference
comes from a galaxy (the point near $R=18.7$).  Some of the scatter
near $R \approx 20$ is probably due to RR~Lyrae variables.  The
unweighted mean $\Delta R$ for objects brighter than $R_{\rm E1} = 21$
is $-0.026 \pm 0.010$, excluding the discrepant points discussed
above.  We attribute this large mean difference to the variable focus
in field E1.

Figure~5(b) plots the magnitude difference $\Delta V \equiv V_{\rm E1} -
V_{\rm C0}$ vs. $V_{\rm E1}$ for the same 190 objects.  This figure
also shows a zero-point offset, though in the opposite direction to
that in Figure~5(a).  The discrepant points have the same cause as those
in the $R$-band comparison.  The unweighted mean $\Delta V$ for objects
brighter than $V_{\rm E1} = 21.5$ is $+0.010 \pm 0.010$, excluding the
most discrepant points.  Again, the magnitude difference depends on
position along the overlap and is most discrepant in the region where
the focus in E1 is the worst.

	Figure~6(a) plots the magnitude difference $\Delta R \equiv
R_{\rm E1} - R_{\rm SE1}$ vs. $R_{\rm E1}$ for the 54 objects common to
the E1 and SE1 fields, matched to within 1 pixel.  A bad column in the
E1 image affects the comparison.  The unweighted mean $\Delta R$ for
objects brighter than $R_{\rm E1} = 22.0$ is $-0.005 \pm 0.015$,
excluding the two most deviant points (produced by objects on the bad
column).  As with the comparison of E1 and C0, the magnitude difference
depends on position within the overlap region -- an effect due to
focus variations in E1.
 
Figure~6(b) plots the magnitude difference for the same 54 objects in the
$V$ band.  The unweighted mean $\Delta V$ for objects brighter than
$V_{\rm E1} = 22.0$ is $-0.013\pm 0.020$, excluding one discrepant
point corresponding to an object on the bad column.

Both sets of comparisons with the photometry of the E1 field show that
the focus variations cause offsets in the zero-points at the level of
a few hundredths of a magnitude.  These offsets do not significantly
affect our conclusions, partly because we remove color offsets between
the data from different fields.  This is discussed in Sec.~\ref{cmd}.
However, the focus variations in E1 do affect our ability to differentiate
between stars and galaxies and, for this reason, we exclude this field
from some of our analyses.

        Figure~7(a) plots the magnitude difference $\Delta R =
R_{\rm S2} - R_{\rm SW2}$ vs. $R_{\rm S2}$ for the 83 objects common to
the S2 and SW2 fields, matched to within 1 pixel.  The discrepant
point near $R_{\rm S2} = 18$ corresponds to a galaxy.  Excluding this
point, the unweighted mean $\Delta R$ for objects brighter than
$R_{\rm S2} = 22.0$ is $+0.007 \pm 0.014$.

Figure~7(b) plots the magnitude difference for the same 83 objects in the
$V$ band.  The unweighted mean $\Delta V$ for objects brighter than
$V_{\rm S2} = 22.0$ is $+0.015\pm 0.012$, excluding the same galaxy and
the next most discrepant point.  The latter is due to an object only
partly within the field.  The good agreement between the photometry from
the S2 and SW2 fields suggests that most of our data has a consistent
zero-point.

The comparison of the photometry from overlapping fields shows that the
zero points of our different fields could differ by approximately
0.01~mag.  The comparison with SVM and CS suggests that the zero points
of their and our studies are also uncertain at the 0.01~mag level.

\subsection {Completeness}
\label{complete}

        To ascertain the completeness of our photometry as a function
of magnitude, we performed artificial star tests for all of the Draco fields.

	In each test, we added 600 artificial stars to both the $R$ and
$V$ frames at random positions.  This number of artificial stars is
10\% or less of the total number of objects in a field.  The added
stars have the profile of the PSF for the given frame.  We reduced the
frames following the same procedure as for the original frames. The
artificial stars are recovered by matching their known input positions
with those of the recovered objects both artificial and real.

	For a given field, we performed six simulations.  In each of the
first three simulations, we added a total of 600 artificial stars to a
frame, 200 for each of the following pairs of $R$ and $V-R$: 22.1 and
0.368; 22.6 and 0.288; and 23.1 and 0.257. The second three simulations
are identical except that they used $R$ and $V-R$ pairs of: 23.35 and
0.258; 23.6 and 0.265; and 23.85 and 0.276. We chose the $R$ and $V-R$
pairs from an isochrone with [Fe/H] = --2.0 and an age of 14~Gyrs taken
from McClure et al. (1987), shifted to the distance and reddening of
Draco given in Webbink (1985) using the relation between $E(B-V)$ and
$E(V-R)$ from Cardelli et al. (1989) (i.e., apparent $(m-M)_{R}= 19.46$
and $E(V-R) = 0.02$).

	Table~5 lists the average recovery percentages of the
artificial stars for each Draco field as a function of $R$. The
uncertainties are the rms scatter of the three simulations. With the
exception of the C0 field, the recovery is almost complete up to
$R$=22.6, with field-to-field variations of about $2.5\%$ or less. The
small loss of artificial stars is caused by mergers with
saturated objects, which ALLFRAME eventually rejects. Of course, an
artificial star could also merge with an object which is not
saturated, though most likely faint. We checked the likelihood of this
by matching the input positions of the artificial stars with the
original positions of the objects from Table~4.  We found that only a few
artificial stars are sufficiently close to the objects for
mergers to occur. Since the artificial star is recovered and the
object it has merged with is probably faint, we decided to include
such recoveries in our recovery percentages.

	For $R=23.1$ and fainter, the completeness percentages are
well below $100\%$ and their field-to-field variations become
substantial.

\section{Color-Magnitude Diagrams}
\label{cmd}

	The objects listed in Tables~4(a) through 4(i) are a
combination of Draco stars, foreground Galactic disk and halo stars,
background galaxies, and a few spurious objects due to bad pixels or
charge overflow columns.  The expected surface density of Draco stars
beyond the tidal boundary is smaller than the total foreground and
background surface density.  Thus, our search for such Draco stars is
more sensitive if we can use additional information to eliminate other
objects from the sample.  This section discusses the two pieces of
information that we use: the morphology of the images and the location
of objects on the color-magnitude diagram.

Two quality-of-fit parameters of DAOPHOT, CHI and SHARP (described in
Sec.~\ref{data}), permit some discrimination between the objects.
Figure~8 is a plot of the $R$--band magnitude vs. SHARP for the C0
field.  Figure~8 shows three distinct groups of points.  First, a
bright, horizontal group with $R \lesssim 16$ whose SHARP values are
widely scattered.  These are saturated stars or spurious objects on
charge overflow columns with CHI values larger than 5.0.  We eliminate
these from all samples discussed subsequently.  Second, a vertical
group whose SHARP values are near zero for objects brighter than $R
\simeq 22$.  For fainter objects, the scatter in the values increases
because of the decreasing signal-to-noise and becomes very wide fainter
than $R \simeq 23$.  Most of the brighter objects in this group are
stars; at fainter levels they become mixed with the third group --
galaxies.  The extended images of galaxies have larger SHARP values
than those of stars, allowing the two groups to be distinguished until
a decreasing signal-to-noise causes the two groups to overlap, as
depicted in Figure~8.  For objects fainter than $R = 21$, exact
differentiation between the two groups based on SHARP values becomes
impossible.

The lines in Figure~8 show the SHARP limits that we have adopted to
separate stars from galaxies.  We set these limits symmetrically about
zero to include nearly all of the objects with negative SHARP values,
which are not contaminated by galaxies.  Thus, ``stars'' must have SHARP
values within 0.25 of zero for $R < 21.0$; within limits increasing
linearly to $\pm$0.50 at $R = 22.0$; and within limits increasing
linearly to $\pm$1.00 at $R = 23.0$.  For objects fainter than $R =
23.0$, we allow any value of SHARP.  In our analysis, we will apply
these limits to the data for all of our fields.

Figure~9 shows color-magnitude diagrams (CMDs, hereafter) plotting $R$
vs. $V-R$ for our nine fields. The $R$-- and $V$--band magnitudes are from
Tables~4(a) through 4(i). We only plot objects with CHI values less
than 5.0 and SHARP values within the limits shown in Figure~8.

The CMD for the C0 field clearly shows the principal sequences of
Draco: in particular, the horizontal branch (HB, hereafter) and red
giant branch (RGB, hereafter).  The distribution of stars on
the HB is somewhat distorted by RR Lyrae stars in Draco with
incorrectly measured colors due to the lack of phase coverage in our
data.  The effect is particularly noticeable for C0 because the $R$--
and $V$--band images are from different nights.

	The CMDs for the E1 and W1 fields also show the HB and RGB, as
expected, since  more than 50\% of both fields are within the measured
tidal boundary of Draco (see Figure~1).

	No easily discernible Draco principal sequences appear on any
of the remaining CMDs. However, this should not be construed as
categorical proof for the absence of Draco stars in these fields.
Instead, it is only certain that these fields have fewer Draco stars
than the C0, E1, and W1 fields.

	Excluding the features due to Draco stars in the CMDs for the
C0, E1, and W1 fields, all of the CMDs display a very similar and
complex morphology produced by Galactic field stars. A sharp blue edge
to the distribution of points is one of the most conspicuous features
seen on the CMDs.  It is due to the main-sequence turnoff of Galactic
field stars smeared in apparent magnitude by the range of distances
along the line of sight (e.g., Reid \& Majewski 1993).  Disk stars
populate mostly the upper part of the blue edge due to the finite
thickness of the disk, whereas the more distant halo stars occupy the
lower part.  Intrinsically fainter main sequence disk stars produce
the band of points extending to the red from the disk turnoff region
to the concentration of points between $V - R = 1.0$ and 1.5, which is
due to the numerous disk stars on the lower main sequence.  The large
number of points at the bottom of each CMD represent both stars and
galaxies not rejected by the increasingly loose SHARP limits required
by the larger photometric uncertainties.

The color of the blue edge should be the same in all of our CMDs.  It
is not.  One possible reason for this is differential reddening due to
interstellar dust. Figure~10 shows the location of our
fields superimposed on a map of the reddening, $E(B-V)$, derived from
dust infrared emission by Schlegel \etal\ (1998).  From this map, we
calculated the average reddening for each field; these are listed in
column two of Table~6.  We adopt an average reddening for each field
instead of correcting each object individually because the reddening
does not vary strongly across our fields and the resolution of the map
is comparable to the size of our fields.  Multiplying the $E(B-V)$
values for each field by 2.673 and 3.315 (Schlegel \etal~1998; the
Landolt values from their Table~6) gives the extinction in the $R$ band,
$A_R$, and $V$ band, $A_V$, respectively. Table~6 lists these values in
columns 3 and 4, respectively.

The fifth column of Table~6 lists the measured color of the
blue edge for all of the fields after correcting for interstellar
reddening and extinction. We estimated these colors by counting the
number of points in a sliding rectangular box in the dereddened CMD.
The box is 0.1~mag wide in color and extends from $R=19$ to $R=14$. As
the box slides from $(V-R)_0=0$ to $(V-R)_0=0.4$ in 0.025~mag steps, the
number of points in the box rapidly increases as it crosses the blue
edge. We normalize the numbers with respect to the highest value and
adopt as the location of the edge that color where the normalized
number is 0.5.

	The C0 field has the bluest edge ($(V-R)_0=0.240$), the W2
field has the reddest ($(V-R)_0=0.340$), and the rest of the fields
have values roughly clustered around a mean of $(V-R)_0=0.304$.  The
difference in the extreme values is visible by eye on the CMDs and is
significantly larger than the 0.007~mag average measurement uncertainty
estimated from a Monte Carlo simulation.  The variation in the edge
colors is likely due to errors in the photometric zero points and is
consistent with the zero-point differences shown in Figures~5 -- 7 for
objects located where fields overlap; see the discussion in
Sec.~\ref{data} for details.

	Figure~11 shows a CMD for the C0 field with no corrections made
for extinction.  The solid lines outline the region where the principal
sequences of Draco are located.  Magnitude $R=16.4$ bounds this region
at the top and $R=23.4$ (the approximate limit to which the photometry
has completenesses greater than 50\%, see Table~5) at the bottom.  The
photometric uncertainty in $V-R$ determines its horizontal width for
objects brighter than $R=22.6$ --- we adopted $\pm 3\sigma_{V-R}$.
Fainter than this limit, the width increases only slightly to keep the
region from including nearly all objects.  The crosses near the
$R=20$~mag level correspond to some confirmed RR~Lyrae stars from Baade
\& Swope (1961), validating the assertion that the distribution of
stars on the HB is somewhat distorted by the variability of RR~Lyrae
stars.  We did not attempt to identify, or to confirm, all of the
possible RR~Lyrae stars in our fields because our data is too sparsely
sampled in time to do so effectively.

	We evaluate and classify every object in every field based on
its CHI and SHARP values and where it is positioned on a CMD.  The
latter evaluation must be done in a consistent photometric system.  We
adopt the photometry of the C0 field, not corrected for extinction, as
this system in this article.  Thus, the $R$--band magnitude of every
object is adjusted by the difference between the $A_R$ for its field
and that for C0.  The $V-R$ of every object is adjusted by the
difference in the reddenings and by the difference in the
reddening-corrected colors of the blue edge.

Objects divide themselves into two categories depending on their
positions in the CMD with respect to the region outlined in Figure~11.
An object is ``in'' if it lies within the region and it is ``out'' if
it lies outside of the region.

\section{Searching for Draco Stars --- Methods and Results}
\label{search}

	In this article we use three methods to search for Draco stars
beyond the measured tidal boundary of the dSph.  1) Plotting
color-magnitude diagrams for combinations of fields and sections of fields
and looking for the principal sequences.  2) Measuring the
field-to-field and 3) intra-field fluctuations in the surface
density of objects selected from the ``in'' and ``out'' regions of the
CMD defined in the previous section.

Each of the above methods makes different assumptions and has its own
strengths and weaknesses.  The first method makes no assumptions about
the distribution of Draco stars on the sky.  However, there must be
enough of these stars in the regions examined to make the principal
sequences visible in the CMD.  The necessity of recognizing the
principal sequences makes this method somewhat subjective.  The other
two methods assume that the Draco stars beyond the tidal boundary have
an irregular distribution on the sky.  The second method searches for
irregularities in the distribution on an angular scale comparable to
the separation between fields, while the third looks for irregularities
on a scale comparable to the angular size of an individual field.
Unlike the first method, these two methods are objective and can
produce quantitative estimates of the number of Draco stars present.
However, they cannot find Draco stars uniformly distributed among all
of the fields --- such as a symmetric halo of uniform density.  We
search for irregularities using the surface density of objects from the
``in'' region of the CMD, $\Sigma_{\rm in}$.  Another weakness of the
second and third methods is that fluctuations in the number of
foreground and background objects can mask those in the Draco stars.
To judge the size of the fluctuations of foreground and background
objects, we examine the surface density of ``out'' objects,
$\Sigma_{\rm out}$.  We will calculate $\Sigma_{in}$ and $\Sigma_{out}$
to many limiting magnitudes.  If no limit is stated, it is
$R = 22.6$, the limit to which our photometry is nearly complete.
We also examine the ratio $\Sigma_{\rm
in}/\Sigma_{\rm out}$, a quantity less sensitive to the correlation
between $\Sigma_{\rm in}$ and $\Sigma_{\rm out}$ caused by the large
contribution of field objects to $\Sigma_{\rm in}$.  We discuss this
complex issue further below.

The following three subsections present our implementation of the
three methods in more detail and describe the results of the search for
Draco stars.  The reader may wish to read the summary of the results in
Section~\ref{discussion} before reading the following subsections.

\subsection{Principal Sequences in the CMD}
\label{cmdsearch}

	If Draco stars are present in a given field, they will lie
close to the principal sequences of Draco in the CMD.  The CMDs for the
fields outside of the tidal boundary of Draco, depicted in Figure~9, do not
show such sequences.  It is especially surprising not to see them in
the CMD for the SE1 field, whose northwest corner is inside of the
measured tidal boundary of  Draco (see Figure~1).  As we remarked above, the
apparent absence of the sequences in a CMD does not mean that there are
zero Draco stars in this field; instead, it poses the question: How
close is the number to zero?

If there are small numbers of Draco stars present in individual fields,
combining the data into a single CMD may increase the visibility of the
principal sequences.  If Draco stars are not uniformly distributed
either within or between fields, it will be more advantageous to
combine just the data from the regions where the Draco stars are most
numerous.  One approach is to select the fields and sections of fields
by trial and error to maximize the visual appearance of the principal
sequences.  Another is to select quadrants of fields having
larger-than-average values of $\Sigma_{\rm in}$ or
$\Sigma_{\rm in}/\Sigma_{\rm out}$.  These are the two approaches
that we take in this article.

	Sections of the E1, SE1, and W1 fields are outside of the tidal
boundary of Draco (see Figure~1). Figure~12(a) is a CMD for the
combined set of objects within these sections.  For comparison,
Figure~12(b) shows the CMD for the C0 field plotted with the same
scale.  These CMDs, and the others that follow, show only those objects
satisfying the CHI and SHARP criteria described in Sec.~\ref{cmd}.  In
addition, all data have been shifted to the photometric system of the
C0 field.  The CMD in Figure~12(a) shows only very weak evidence for
the presence of Draco stars: a small clump of points in the HB region
($R \approx 20$, $V - R \approx 0.25$) and a faint lower RGB.  Both
features could occur by chance due to a small number of points in those
regions of the CMD.  However, Figure~12(c) shows that adding the
objects from the E2 and W2 fields to the data in Figure~12(a) makes the
principal sequences visually more prominent.  In contrast, Figure~12(d)
shows that adding the objects from the N2, S2, and SW2 fields to the
data in Figure~12(a) does not have the same effect.

Admittedly, the presence or absence of the principal sequences of Draco
in Figures~12(a), (c), and (d) is subjective.  The human eye is better
suited to discerning very weak structural features in gray-scale images
than on a scatter plot, especially if the number of points is small.
Thus, to make the decision more objective, we replot Figure~12 as
gray-scale Hess diagrams in Figure~13.  We calculated the density of
points in these diagrams on a $150 \times 150$ pixel grid using a
fixed-width kernel of the form $(1-(\Delta R/w_R)^2-(\Delta
(V-R)/w_{V-R})^2)^2$.  Here $\Delta R$ and $\Delta (V-R)$ are the
distances of the point from the center of a pixel in $R$ and $V-R$,
respectively, and $w_R$ and $w_{V-R}$ are two constants that determine
the width of the kernel.  For Figure~13(a), $w_R = 0.075$ and $w_{V-R} =
0.30$, while for Figures~13(b), (c), and (d) $w_R = 0.062$ and $w_{V-R}
= 0.25$.  In Figure~13, white corresponds to zero density and black to
35\% of the peak density.  The contour in Figures~13(a), (c), and (d)
is an isopleth from Figure~13(b) and outlines the principal sequences
of Draco.  Figures~13(a) and (c) show density enhancements at the
expected locations of the HB and, less prominently, the lower RGB.  In
contrast, Figure~13(d) does not.

Panels (a) and (c) of Figures~12 and 13 imply that there are Draco
stars beyond the tidal boundary measured by IH.  Panels (c) and (d) of
Figures~12 and 13 show that fewer, possibly no, Draco stars are present
in the N2, S2, and SW2 fields, implying an irregular azimuthal
distribution for the Draco stars beyond its measured tidal boundary.
If there are Draco stars in the E2 and W2 fields, as Figures~12(c) and
13(c) suggest, then these stars are at least $60$~arcmin away from the
center of Draco approximately along the extension of its major axis.

Taking our alternative approach, we select quadrants from our fields
that are entirely beyond the measured tidal boundary and have values of
$\Sigma_{\rm in}$ and $\Sigma_{\rm in}/\Sigma_{\rm out}$ larger than
the average.  This procedure excludes all quadrants in the C0, E1, and
W1 fields and the northwest quadrant of the SE1 field.  If high values
of $\Sigma_{\rm in}$ and $\Sigma_{\rm in}/\Sigma_{\rm out}$ are
indicative of the presence of Draco stars, the CMD for the data from
these regions should show the principal sequences.  Figure~14(a) is the
CMD for the quadrants with higher than average $\Sigma_{\rm in}$: SE1 -
northeast; N2 - northwest; E2 - northeast, southwest, and southeast;
SW2 - northeast, southwest, and southeast; W2 - northeast and
southwest.  Figure~14(c) is its corresponding gray-scale Hess diagram,
calculated as for Figure~13 with kernel widths of $w_R = 0.062$ and
$w_{V-R} = 0.25$.  Figure~14(b) is the CMD for the quadrants with
higher than average $\Sigma_{\rm in}/\Sigma_{\rm out}$: SE1 -
northeast; N2 - northeast and northwest; E2 - northeast and southwest;
SW2 - northeast; W2 - northeast, northwest, and southwest.
Figure~14(d) is its gray-scale Hess diagram.  The contour in both
Figures~14(c) and (d) is the same as that in Figure~13(a).  None of
these figures shows the principal sequences as clearly as Figures~12(c)
and 13(c).  There are clumps of points in the general vicinity expected
for the HB, but Figures~14(c) and (d) show that these are brighter than
expected for Draco HB stars.  Thus, either the number of Draco stars in
the quadrants entirely beyond the tidal boundary is too small, or the
fluctuations in the number of field objects in those quadrants is too
large, for $\Sigma_{\rm in}$ or $\Sigma_{\rm in}/\Sigma_{\rm out}$ to
be a good indicator of the presence of Draco stars.

We have tried many other combinations of quadrants and entire fields,
both with and without the regions of E1 and W1 outside of the measured
tidal boundary, and none show the principal sequences more clearly than
Figures~12(c) and 13(c).

In conclusion, we have found evidence for Draco stars beyond the tidal
boundary measured by IH.  The distribution of these stars is uncertain,
but they may extend as far as 1.3~degrees from
the center of Draco.

\subsection{Surface Density Fluctuations}
\label{fluctsearch}

	If the surface density of Draco stars beyond the measured tidal
boundary is non-uniform on a scale comparable to the angular separation
between the fields, then the value of $\Sigma_{\rm in}$ should
fluctuate from field to field.  If these fluctuations are due to Draco
stars, then the field-to-field fractional rms scatter around the mean
$\Sigma_{\rm in}$ should increase with the limiting magnitude of the
sample because of the rapid increase in the number of stars along the
RGB.  However, field objects, whose projected distribution can
be non-uniform too, make the largest contribution to $\Sigma_{\rm
in}$.  Therefore, it is imperative to check for other causes for the
fluctuations in $\Sigma_{\rm in}$.  To help with this, we measure the
fractional rms scatter around the mean $\Sigma_{\rm out}$ and,
additionally, around the mean $\Sigma_{\rm in}/\Sigma_{\rm out}$, both
as a function of limiting magnitude.  However, we caution that it is
possible that the mixture of disk stars, halo stars, and background
galaxies that contribute to $\Sigma_{\rm in}$ could fluctuate
differently from those that contribute to $\Sigma_{\rm out}$.  Despite this,
significant fluctuations in $\Sigma_{\rm out}$ would suggest the potential
for fluctuations in $\Sigma_{\rm in}$ due to field objects.

Galaxies are clustered on the sky and thus contribute to the
fluctuations in both $\Sigma_{\rm in}$ and $\Sigma_{\rm out}$.  We can
partially exclude galaxies from our data sets based on how well an
object matches the stellar PSF, as discussed in Sec.~\ref{cmd}.  If
this exclusion significantly reduces the size of the fluctuations, it
would suggest that there is a potential for galaxies to still cause
fluctuations in $\Sigma_{\rm in}$, since we know that not all galaxies
can be eliminated.  This problem is greatest for the faintest objects,
whose low signal-to-noise ratio makes galaxy-star separation impossible.

The panels in Figure~15 plot the fractional rms scatter around the mean
of $\Sigma_{\rm in}$, $\Sigma_{\rm out}$, and $\Sigma_{\rm
in}/\Sigma_{\rm out}$ measured in seven regions as a function of
limiting magnitude.  These are shown as solid points and should be
compared with the scatter expected from Poisson noise, shown as open
squares.  The seven regions are: SE1 (excluding the small section
within the measured tidal boundary), W1 (also excluding the section
within the tidal boundary), N2, E2, S2, SW2, and W2.  We exclude the
entire E1 field from this figure because the large focus variations
seriously degrade the galaxy-star separation.  However, we did not
exclude E1 in the discussion of CMDs because the increase in the
contamination by galaxies does not significantly affect the detection
of the principal sequences.  The error bar shown for the measured
fractional rms scatter is the sampling uncertainty for a sample of
seven measurements.  To indicate the contribution of galaxies to the
rms scatter, the left-hand panels use all objects, regardless of their
SHARP and CHI values, while the right-hand panels exclude objects which
either have CHI values larger than 5.0 or SHARP values outside of the
limits shown in Figure~8.

Figure~15(a) plots the fractional rms scatter in $\Sigma_{\rm in}$ as a
function of limiting magnitude for all objects, regardless of their
SHARP and CHI values.  The figure shows that the field-to-field
fluctuations in $\Sigma_{\rm in}$ are larger than those expected from
Poisson noise for all limiting magnitudes.  Figure~15(b) is the
corresponding plot for $\Sigma_{\rm out}$, though note the different
vertical scale.  The field-to-field fluctuations in $\Sigma_{\rm out}$
are also larger than those expected from Poisson noise, though
generally smaller than those for $\Sigma_{\rm in}$.  Around $R=22.2$,
the fluctuations in $\Sigma_{\rm out}$ begin to increase with limiting
magnitude and exceed those in $\Sigma_{\rm in}$.  This increase is due
to the progressively larger contribution from galaxies with increasing
magnitude; a particularly large contribution comes from the SW2 field,
which contains an x-ray galaxy cluster (Zwicky 8197; Ebeling
\etal\ 1998).  Replotting the figure excluding the SW2
field reduces the fluctuations significantly, though they still
generally remain larger than Poisson noise.  Unequal completeness for
the different fields could also contribute to the increase in the
fluctuations for the faintest limiting magnitudes, though its contribution is
probably small since Figure~15(a) does not show a comparable increase.

Figure~15(c) plots the fractional rms scatter in
$\Sigma_{\rm in}/\Sigma_{\rm out}$ as a function of limiting magnitude.
The pluses are the values expected for the observed fractional rms scatter
if the scatters in $\Sigma_{\rm in}$ and $\Sigma_{\rm out}$ are
uncorrelated, i.e., they add in quadrature.  All of the observed points, the
solid squares, are below the pluses.  This result shows that fluctuations in
$\Sigma_{\rm in}$ and $\Sigma_{\rm out}$ are correlated.  We conclude that
the fluctuations in $\Sigma_{\rm in}$ seen in Figure~15(a) are not solely
attributable to Draco stars beyond the measured tidal boundary.
Instead, field stars and galaxies must contribute to these
fluctuations.  We can partially remove the contribution from galaxies
by excluding objects based on their values of CHI and SHARP; however,
there is no similar way to remove the contribution from field stars.
The right-hand panels of Figure~15 show the corresponding plots with
the galaxies partially removed.

Figure~15(e) shows that fluctuations in $\Sigma_{\rm out}$ are
comparable with Poisson noise up to about $R=22.4$.  At fainter limiting
magnitudes, the fluctuations begin to increase as in Figure~15(b) --- a
reflection of looser SHARP limits (see Figure~8) and the galaxy cluster
in the SW2 field.  Again, excluding the data from the SW2 field reduces
this increase.  Comparing Figures~15(b) and (e) shows that removing
galaxies from the data has reduced the field-to-field fluctuations in
$\Sigma_{\rm out}$ for all limiting magnitudes and, for $R < 22.4$,
reduces it to the level of Poisson noise.  Based on this comparison,
field stars do not cause significant fluctuations in $\Sigma_{\rm
out}$, suggesting that they are not likely to cause them in
$\Sigma_{\rm in}$ either.

Comparing Figures~15(a) and (d) shows that removing galaxies from the
data has reduced the field-to-field fluctuations in $\Sigma_{\rm in}$
for all limiting magnitudes, but not to the level of Poisson
noise.  This contrast with Figures~15(b) and (e) argues that significant
fluctuations remain, most likely due to the presence of Draco stars
beyond the measured tidal boundary.  Note that the points in Figure~15
are not independent; hence, the statistical significance of the
deviation from Poisson noise must be evaluated at only a few points.
Replotting Figure~15(d) excluding the objects from the W1 field shows
fluctuations that are comparable to the expected Poisson noise, arguing
that the objects from the section of the W1 field beyond the measured
tidal boundary have the greatest contribution to the fluctuations seen
in Figure~15(d).  The fluctuations in the number of Draco stars in the
more distant fields cannot be larger than those expected from Poisson
noise, though they might still be present.  We conclude that there are
Draco stars in the region of the W1 field beyond the tidal boundary
measured by IH.  However, we do not find
clear evidence from the fluctuations of $\Sigma_{\rm in}$ to support
the suggestion from the CMDs in Sec~\ref{cmdsearch} for Draco stars at
larger radii.

Figures~15(d) and (f) show that the fluctuations in $\Sigma_{\rm
in}/\Sigma_{\rm out}$ are closer to those expected from Poisson noise
than the fluctuations in $\Sigma_{\rm in}$.  This is probably due to
the increased Poisson noise contributed by $\Sigma_{\rm out}$, which
dilutes the signal of Draco stars beyond the tidal boundary from
$\Sigma_{\rm in}$.  Figure~15(f) shows that removing galaxies from the
data has reduced the correlation between fluctuations in $\Sigma_{\rm
in}$ and $\Sigma_{\rm out}$ for all limiting magnitudes.  Only fainter
than a limiting magnitude of $R = 22.6$ does the correlation become
significant.  The likely causes of this are unequal completeness
between fields and faint galaxies whose SHARP values are within the
loosening limits shown in Figure~8.

Figure~15 shows that galaxies produce significant and correlated
fluctuations in  $\Sigma_{\rm in}$ and $\Sigma_{\rm out}$.  Eliminating
objects based on their SHARP and CHI values reduces these correlations
to statistically insignificant levels for all limiting magnitudes for
which our data are complete.  Significant fluctuations remain only in
$\Sigma_{\rm in}$ for limiting magnitudes brighter than $R = 22.6$.  We
attribute these to the presence of Draco stars beyond the measured
tidal boundary, particularly in the W1 field.

Table~7 and Figure~16 provide information on $\Sigma_{\rm in}$,
$\Sigma_{\rm out}$, and $\Sigma_{\rm in}/\Sigma_{\rm out}$ for each
field at the limiting magnitude for which our photometry is complete,
$R = 22.6$.  Table~7 has three blocks: the first lists the surface area
of the field in square arcmin; the second lists the values of the two
surface densities in objects per square arcmin and their ratio for all
objects, regardless of their SHARP and CHI values; the third block is
the same as the second for objects with CHI values less than 5.0 and
SHARP values within the limits depicted in Figure~8.  We list two
uncertainties below each surface density or ratio entry: the internal
uncertainty, $\sigma_{\rm int}$, and the external uncertainty,
$\sigma_{\rm ext}$.  The first is the Poisson noise and the second is
the the estimate based on the rms scatter around the mean value for the
four quadrants of the field.  The last three columns of the table give
1) the mean value of the two surface densities or their ratio and the
rms scatter around this mean calculated for the sets of fields
described in the notes to the table;  2) the two $\chi^{2}$ values for
the scatter around each mean calculated with $\sigma_{\rm int}$ and
$\sigma_{\rm ext}$; and 3) the probability of exceeding each $\chi^{2}$
value by chance.

The top three panels of Figure~16 plot, from left to right, $\Sigma_{\rm
in}$, $\Sigma_{\rm out}$, and $\Sigma_{\rm in}/\Sigma_{\rm out}$ for a
selection of fields from the second block of Table~7.  The middle three
panels are the same plots for data from the third block of Table~7.  In
these six panels, the error bars on all of the points are $\sigma_{\rm
int}$.  The bottom three panels are the same as the middle panels,
except that the error bars are $\sigma_{\rm ext}$.  For each column of panels,
the ordinates have the same fractional range of values to simplify the
visual comparison of fractional scatters.  We exclude from the
figure fields C0 and E1.  C0 falls entirely within the tidal boundary
and also contains an unusually large number of spurious objects which
compromise $\Sigma_{\rm out}$.  E1 has a variable focus which
compromises $\Sigma_{\rm in}$ and $\Sigma_{\rm out}$.  The figure
includes the section of field W1 beyond the tidal boundary, denoted as
W1$^*$ on the abscissa.  The values for this section are not in
Table~7; they are for the top panels: $\Sigma_{\rm in} = 2.79\pm
0.12$, $\Sigma_{\rm out} = 4.53\pm 0.15$, and $\Sigma_{\rm
in}/\Sigma_{\rm out} = 0.615\pm 0.032$ and for the middle panels
$\Sigma_{\rm in} = 1.902\pm 0.095$, $\Sigma_{\rm out} = 3.19\pm 0.12$,
and $\Sigma_{\rm in}/\Sigma_{\rm out} = 0.596\pm 0.038$.  Figure~16
plots values for the entire SE1 field because the portion within the
tidal boundary is small --- excluding this portion has a negligible effect
on the values.

The top three panels in Figure~16 show the large scatter among fields
that causes the values of the fractional rms to be larger than that
expected from Poisson noise at $R=22.6$ in the left-hand panels of
Figure~15.  The high value of $\Sigma_{\rm out}$ for the SW2 field is due
to galaxies from the cluster Zwicky 8197.  This cluster extends into
the S2 field, also raising its $\Sigma_{\rm out}$.  Eliminating objects
based on their SHARP and CHI values reduces the scatter among fields as
shown in the middle row of panels.  This is reflected in the reduction
of the $\chi^2$ values calculated with $\sigma_{\rm int}$ between blocks
two and three of Table~7.  In the middle row of panels, the W1$^*$
field has the highest value of $\Sigma_{\rm in}$ and of $\Sigma_{\rm
in}/\Sigma_{\rm out}$ and a value of $\Sigma_{\rm out}$ comparable to
those for the other fields.  This demonstrates that this region
contains Draco stars.

Excluding W1$^*$, the middle row of panels still shows more scatter
than expected from Poisson noise and this is confirmed by the $\chi^2$
values and their probabilities in the third block of Table~7.  Is this
evidence for Draco stars in these fields?  It is at best weak evidence,
since this scatter is larger for $\Sigma_{\rm out}$ than for
$\Sigma_{\rm in}$ and the scatter in $\Sigma_{\rm in}$ has a 9\%
probability of being caused by Poisson noise.  Though the scatter in
$\Sigma_{\rm in}$ could be caused by Draco stars, we judge that this is
unlikely because of the significant fluctuations in $\Sigma_{\rm out}$,
which cannot be caused by Draco stars.  Other possible sources for
fluctuations in either surface density are differences in image quality
that affect the rejection of objects based on their CHI and SHARP values
and the clustering of faint galaxies that cannot be rejected because of
their low signal-to-noise ratio.

The bottom row of panels in Figure~16 and the $\chi^2$ values calculated
with $\sigma_{\rm ext}$ in Table~7 test if the fluctuations among fields
are caused by the clustering of galaxies or field stars on scales
comparable to the size of a single field.  These $\chi^2$ values in the
second block of Table~7 show that there are significant fluctuations
within fields for all objects without regard to their CHI and SHARP
values.  In contrast, these $\chi^2$ values in the third block are not
significantly different from those calculated with $\sigma_{\rm int}$ ---
the scatter of the surface densities and their ratio shown in the
bottom panels of Figure~16 are still larger than the error bars.  We
conclude that this scatter is due to differences among fields, either
in the surface densities of galaxies or stars or in the ability to
reject objects on the basis of their CHI and SHARP values.
Unfortunately, we cannot quantify these different contributions and so
we can only place an upper limit on the fluctuations due to Draco stars
--- the rms scatter around the mean $\Sigma_{\rm in}$ given in the
third block of Table~7.

\subsection{Intra-field Fluctuations}

	If the Draco stars beyond the measured tidal boundary have a
non-uniform distribution on an angular scale comparable to that of a
single field, then $\sigma_{\rm ext}$ should be larger than
$\sigma_{\rm int}$ for $\Sigma_{\rm in}$ and $\Sigma_{\rm
in}/\Sigma_{\rm out}$ while comparable for $\Sigma_{\rm out}$.  Blocks
2 and 3 in Table~7 list these uncertainties.  The presence of Draco
stars in the C0, E1, and W1 fields is evident in Table~7, which shows
that $\sigma_{\rm ext}$ is larger than $\sigma_{\rm int}$ in the
second and third blocks.  The difference in size is largest for
$\Sigma_{\rm in}$ and $\Sigma_{\rm in}/\Sigma_{\rm out}$.  While $\sigma_{\rm ext}$ is larger than $\sigma_{\rm int}$ for $\Sigma_{\rm out}$, the
difference is only statistically significant for E1, which has a PSF that
varies strongly across the field (see Sec.~\ref{data}).
 
The relative sizes of the values for $\sigma_{\rm int}$ and
$\sigma_{\rm ext}$ listed in the second block of Table~7 show that
there are significant fluctuations in both surface densities and in
their ratio on the scale of a single field.  However, except for the
fields C0, E1, and W1, the values in the third block of Table~7 show no
statistically significant differences.  Thus, we attribute the larger
$\sigma_{\rm ext}$ in the second block to the presence of galaxies
(note the large values of $\sigma_{ext}$ for the SW2 field) and
conclude that comparing $\sigma_{\rm int}$ and $\sigma_{\rm ext}$
provides no evidence for Draco stars beyond the measured tidal
boundary.

\section{Summary and Discussion}
\label{discussion}

The presence of principal sequences in the CMD (Figures 12~(a) and
13(a)) for the regions of the E1, W1, and SE1 fields beyond the tidal
boundary measured by IH demonstrate the presence of Draco stars in at
least some of these regions.  The value of $\Sigma_{\rm in}$ for the
region of W1 beyond the measured tidal boundary confirms that Draco
stars are in this region.  We think that the CMDs shown in
Figures~12(c) and 13(c) provide evidence (primarily in the visibility
of the HB) that the Draco stars extend into the more distant fields E2
and W2.  However, we acknowledge that others may disagree because the
principal sequences are present only at the level of the noise.  The
field-to-field scatter in the values of $\Sigma_{\rm in}$ for the more
distant fields is comparable to that of $\Sigma_{\rm out}$ and so the
values of $\Sigma_{\rm in}$ can neither prove nor disprove the presence
of Draco stars at these larger radii.  A confirmation of the membership
of these stars must come from other techniques, such as measurements of
radial velocity, proper motion, metallicity, or luminosity.

If there are Draco stars in our E2 and W2 fields, this implies that
the surface density of Draco decreases with increasing
radius much more slowly beyond a radius of about 30~arcmin than
between 10 and 20~arcmin.  Models of the tidal stripping of stars from
a dSph produce such a break in the profile at the approximate radius
where the stars change from being mostly bound to mostly unbound
(e.g., Johnston et al. 1999).  If this interpretation is correct, the
stars at large radii should have a mean velocity above that of Draco
on one side and below on the other (Oh \etal\ 1995; Piatek \& Pryor
1995; Klessen \& Kroupa 1998; Johnston et al. 1999).  Until such
measurements exist, alternative explanations are also viable.  For
example, the creation of a spatially extended population of bound
stars either during the formation of the dSph or later when gas was
driven to large radii by stellar winds or supernovae.  Gas is seen in
large shells around some dwarf irregular galaxies (Puche \& Westpfahl
1994).

The greater visibility of the principal sequences in the CMDs of
Figures~12(c) and 13(c) compared to those of Figures~12(d) and 13(d)
indicates that any Draco stars present at large radii concentrate
mainly to the east and west of the galaxy.  This suggests that the
stars beyond the measured tidal boundary are along extensions of the
major axis, but demonstrating this requires data with more complete
azimuthal coverage.

We find that searching for Draco stars through fluctuations in the
surface densities of objects in fields where the expected number of
these stars is small compared to the numbers of foreground field stars and
background galaxies is subject to significant uncertainties.  The
left-hand panels of Figure~15 show that galaxies produce fluctuations
larger than expected from Poisson noise over a broad range of limiting
magnitudes, thus potentially masking fluctuations due to Draco stars.
Removing objects based on the morphology of their images can greatly
reduce these fluctuations (see the right-hand panels of Figure~15).
However, this procedure cannot be completely effective, particularly when
applied to objects with low signal-to-noise ratios, and can introduce
spurious fluctuations if the focus or seeing varies among fields.

We estimate the surface density of Draco stars beyond the measured
tidal boundary using the $\Sigma_{in}$ values from the third block of
Table~7 and the equivalent value for the W1 field given in the text.
The most reliable estimate is for this region of the W1 field.  The
difference between the $\Sigma_{in}$ for this region and the mean value
for the SE1, N2, E2, S2, SW2, and S2 fields, given in Table~7, yields
$0.30 \pm 0.10$~stars~arcmin$^{-2}$, while using the lowest
$\Sigma_{in}$, that for the SE1 field, instead of the mean yields $0.39
\pm 0.11$~stars~arcmin$^{-2}$.  The comparison of these values with
those of IH and with theoretical models requires that they be expressed
as a fraction of the central surface density of Draco.

To calculate a central surface density of Draco, we derived its density
profile along the major axis in the C0 field by counting in elliptical
annuli those objects that contribute to the $\Sigma_{in}$ listed in the
third block of Table~7.  The ellipticity and position angle of these
annuli are from IH.  The greater incompleteness of the C0 field (see
Table~5) was approximately corrected by multiplying the number of
objects between $R = 22.1$ and 22.6 (38\% of the total number) by
0.95/0.80 = 1.19.  The resulting profile agrees well with that of IH
and yields a central surface density of $20.0 \pm
1.4$~stars~arcmin$^{-2}$.  The Poisson uncertainties in the binned
densities account for most of the uncertainty in this value.  We will
present the radial density profile for Draco based on the data from all
of our fields elsewhere (Piatek \etal\ 2000).

Our estimate of the surface density of Draco stars in the W1 field
beyond the measured tidal boundary is thus $0.015 \pm 0.005$ to $0.020
\pm 0.006$ of the central density.  This estimate is within the range
of values at this radius shown in Figure~2 of IH.

An estimate of the surface density of Draco stars in a field at larger
radii is the difference between its value of $\Sigma_{in}$ in the third
block of Table~7 and the lowest value of $\Sigma_{in}$, that for the S2
field --- which is assumed to be the background.  The largest
difference, for the SW2 field, is $0.18 \pm 0.08$~stars~arcmin$^{-2}$
or $0.0090 \pm 0.0039$ of the central surface density.  We emphasize
that this estimate is suspect both because the SW2 field contains a
rich cluster of galaxies which contaminates $\Sigma_{in}$ and because
the differences are not statistically significant (see the discussion in
Sec.~\ref{fluctsearch}).  The second largest difference is for the E2
field: $0.17 \pm 0.08$~stars~arcmin$^{-2}$ or $0.0083 \pm 0.0039$ of
the central surface density.  These values are consistent with the
lowest point of the profile shown in Figure~2 of IH (though this is
only the lowest point that is above their assumed background).  If
there are Draco stars in the E2 and W2 fields and not in the N2, S2,
and SW2 fields, as suggested by Figures~12(c) and (d) and 13(c) and
(d), then the above value for the E2 field is probably an upper limit
on their surface density.  However, we again emphasize that the
evidence for field-to-field fluctuations in the surface density of
Draco stars at large radii is weak (see Sec.~\ref{fluctsearch}).

	 Several other groups have reported discoveries of stars near or
beyond the tidal boundary of the Sextans, Carina, and Sagittarius dSphs.

Gould \etal\ (1992) identified at least five faint metal-poor dwarf
stars, apparently all at a common distance of about 100~kpc, in their
CT1 field.  This field is about 100~arcmin away from the center of
Sextans along the major axis.  The properties of these stars are
consistent with membership in Sextans and Gould \etal\ (1992) estimate
a surface density, based on these stars, of about 0.01 of the central
value for Sextans.  The IH value for the major-axis tidal radius of
Sextans is $160 \pm 50$~arcmin, implying that the five stars are within
the tidal boundary.  The surface density found by Gould \etal\ (1992)
is consistent with the IH profile for Sextans.

Kuhn et al.\ (1996; hereafter KSH) found evidence for Carina stars in
fields up to $2^{\circ}$ away from the center of this galaxy along the
extension of its major axis.  Their approach is similar to our method
of searching for field-to-field fluctuations in $\Sigma_{\rm in}$,
except that it subtracts an estimate of the contribution from
non-member objects.  They use least squares to estimate the
contribution to the CMD of each field from both the CMD of a background
field, containing non-member stars and galaxies, and a CMD which is the
difference between the CMD of a field centered on Carina and the
background CMD, intended to contain only Carina stars.  Out of the
eight fields along the extension of the major axis and not centered on
Carina, seven have surface densities of Carina stars larger than those
of the two fields 1$^\circ$ on either side of the center along the
extension of the minor axis, which were taken to represent the
background.  The two nearest of the eight fields have the highest
measured contribution from the Carina CMD.  The center of the nearest
field is slightly within the tidal boundary measured by IH, so the
presence of Carina stars in this field is expected and, as KSH note, is
consistent with the profile measured by IH.  The second nearest field
is just outside of the measured tidal boundary of Carina, so the
inferred presence of Carina stars there is reminiscent of our evidence
for Draco stars in the regions of the E1 and W1 fields outside of the
tidal boundary of Draco.

Our values for $\Sigma_{\rm in}$ for fields about 1$^\circ$ from the
center of Draco, shown in the left-most panel of the bottom row of
Figure~16, display less scatter than do the values for the contribution
of the Carina CMD to the CMDs of the KSH fields. This is the case even
after excluding the two innermost of their fields.  The scatter is
closer to what we see before removing galaxies (see the left-most panel
of the top row of Figure~16).  KSH do not remove galaxies from their
sample.  The uncertainty for the Carina contribution estimated by KSH
comes from the scatter of the values from the four quadrants of the
field, as do our error bars in the bottom row of Figure~16.  Their
uncertainties vary more between fields than do ours.  This may be due
to either or both of the following:  the clustering of galaxies on
angular scales comparable to the size of the field and the small
numbers of objects in each bin of their least-squares fit.  Because of
the above concerns about galaxies and the variable uncertainties, we
think that evidence for Carina stars 1$^\circ$ or more from the center
is weaker than KSH claim and is probably no stronger than our evidence
for Draco stars in the E2 and W2 fields.

Majewski~\etal\ (2000; hereafter M2K) identified a ``substantial
extratidal population from Carina'' using Washington-band and
DDO51-band photometry.  With this technique it is possible to
distinguish metal-poor giants from the more numerous disk and halo
dwarf stars (Geisler 1984) to a degree that depends on photometric
accuracy.  Based on the size of this population beyond the tidal
boundary measured by IH, M2K estimated that Carina loses about 27\% of
its mass per Gyr.  Spectroscopy of three bright giant candidates
outside of the tidal boundary confirmed that they have velocities
consistent with that of Carina.  Notwithstanding this confirmation,
Morrison~\etal\ (2000) argued that photometric errors of 0.1~mag, the
limit adopted by M2K, cause dwarfs from the Galactic disk and halo to
contaminate significantly the population of fainter candidate giants.
The photometric and spectroscopic data of M2K demonstrate that there
are Carina stars beyond its measured tidal boundary.  However, the
number of these stars and their distribution in azimuth and radius
remain uncertain because of the incomplete areal coverage and variable
limiting magnitudes of the M2K fields, and the concerns raised by
Morrison~\etal\ (2000).

Since the discovery of the Sagittarius dSph (Ibata \etal\ 1994), stars
associated with this galaxy have been found at ever increasing angular
distances from its center.  Mateo \etal\ (1996) and Fahlman
\etal\ (1996) found main-sequence stars of Sagittarius about 10$^\circ$
from the center.  Mateo \etal\ (1996) also found RR~Lyrae stars
associated with Sagittarius in their field, as did Alard (1996) and
Alcock \etal\ (1997) in fields 8$^\circ$ to 11$^\circ$ from the center
on the opposite side.  Ibata \etal\ (1997)  detected main-sequence
stars in a $22^{\circ} \times 8^{\circ}$ region on the side of
Sagittarius away from the Galactic plane.  They infer a minor axis
limiting radius of 4$^\circ$ and a major to minor axis ratio of 3:1,
implying a major axis limiting radius of 12$^\circ$.  However, Mateo
\etal\ (1998) detected main-sequence stars over the range of 10$^\circ$
to 34$^\circ$ from the center of Sagittarius.  Their measured surface
density profile along the major axis has a break at a radius of about
20$^\circ$.  Majewski \etal\ (1999) report a possible detection of
Sagittarius stars at similar angular distances.  Recently, a survey for
carbon stars in the Galactic halo (Totten \& Irwin 1998; Totten \etal\ 2000)
has found giant carbon stars distributed on a great circle passing through
the position of Sagittarius.  Ibata \etal\ (2000a) interpret these stars
as a stream of tidal debris extending completely around the orbit of
Sagittarius.  Ibata \etal\ (2000b) also argue that halo substructure
found in the Sloan Digital Sky Survey (Yanny \etal\ 2000; Ivezi\'{c}
\etal\ 2000) is tidal debris from Sagittarius.

The existence of stars beyond the tidal boundary of Sagittarius is
established and they are very likely due to tidal stripping.  Is this
the case for the Galactic dSphs at larger galactocentric distance, for
which tides are currently less important?  Our study and others,
particularly that of Majewski \etal\ (2000) for Carina, show that
several dSphs extend beyond the tidal boundaries measured by IH.
However, these stars have not been found to extend to as large a
distance along the major axis as for Sagittarius, so alternative
explanations remain viable.  The most mundane explanation is that
these stars are within the tidal boundary of their dSph.  This
boundary is uncertain because there is no fundamental reason why a
King model (King 1966) should correctly describe a dSph at all radii.
For example, a dSph could be surrounded by an extended population of
bound stars formed when gas in the galaxy was expelled by supernovae
and stellar winds.

Convincing evidence for tidal stripping would be the existence of a
very extended distribution of stars along some axis, particularly if
supported by the presence of a velocity gradient in these stars (Oh
\etal\ 1995; Piatek \& Pryor 1995; Johnston \etal\ 1995).  Though this
axis must be along the orbit, it is not clear that this must be the
major axis of the bound stars since the galaxy might have formed with
the major axis not in the orbital plane.  Indeed, an alignment between
these two could argue for a strong tidal interaction, though this might
also be a result of the dSph forming in the vicinity of the Galaxy.
Extensions of the work presented in this paper that could provide
better evidence for tidal stripping are deeper two-color photometry
with good seeing -- preferably followed up with radial velocity or
proper motion measurements, searches for RR~Lyrae variables, and three-color
photometry that is able to distinguish metal-poor giants from disk and
halo dwarfs.

\acknowledgments

We thank Janie Fultz for proofreading the text and Mike Irwin for
sending us the image of the IH stellar surface density in the vicinity
of Draco included in Figure~2.  The research of CP and SP was
supported by NSF grant AST 96-19510, while that of EWO was partially
supported by NSF grant AST 96-19524.  This research has made use of the
Astronomical Data Center (ADC) at the NASA Goddard Space Flight
Center.

\clearpage
\begin{center}
Figure Captions
\end{center}

Fig. 1. The locations of our fields superimposed on the Digitized Sky
Survey image centered on Draco. The ellipse represents the measured
tidal boundary of Draco.

Fig. 2. The locations of our fields superimposed on a map of the surface
density of stars and galaxies derived by Irwin \& Hatzidimitriou (1995).

Fig. 3. A comparison of the V-band magnitudes derived by Stetson et~al.
(1985) ($V_{\rm SVM}$) and the authors ($V_{\rm us}$) for the 95 objects
common to both studies. All of the objects are in our C0 field.

Fig. 4. A comparison of the V-band magnitudes derived by Carney \&
Seitzer (1986; CS) ($V_{\rm CS}$) and the authors ($V_{\rm us}$).  CS
measured two independent magnitudes for some objects and these appear
twice among the 1382 points in the plot.  A point marked with a slanted
cross corresponds to a confirmed RR Lyrae star.

Fig. 5. (a) A comparison of the R-band magnitudes for the 190
objects common to the E1 and C0 fields. (b) The same as Fig.~5a for
the V band.

Fig. 6. (a) A comparison of the R-band magnitudes for the 54 objects
common to the E1 and SE1 fields.  (b) The same as Fig.~6a for the V
band.

Fig. 7. (a) A comparison of the R-band magnitudes for the 83 objects
common to the S2 and SW2 fields.  (b) The same as Fig.~7a for the V
band.

Fig. 8. R-band magnitude vs. SHARP for all of the objects in the C0
field. The lines represent the SHARP limits that we have adopted in this
study.

Fig. 9. Color-magnitude diagrams for our 9 fields. The $R$ and $V$
magnitudes are from Tables~4(a) through 4(i). Only objects with CHI
values less than 5 and SHARP values within the limits shown in Fig.~8
are shown on these diagrams.  The principal sequences of Draco are
apparent in the diagrams for the C0, E1, and W1 fields.

Fig. 10. The location of our fields superimposed on the map of
$E(B-V)$ derived from dust infrared emission by Schlegel \etal\
(1998).  Lighter shades correspond to larger reddening.

Fig. 11.  Color-magnitude diagram for the C0 field. The $R$ and $V$
magnitudes are not corrected for reddening and extinction.  The solid
contour encloses our adopted region containing stars likely to belong
to Draco.  Its width is approximately determined by the color uncertainty
at each magnitude.  The few points marked with a cross correspond to some
confirmed RR~Lyrae stars.

Fig. 12. (a) Color-magnitude diagram for the combination of those
sections of the E1, SE1, and W1 fields outside of the measured tidal
boundary of Draco.  (b) Color-magnitude diagram for the C0 field. This
diagram serves as a reference. (c) The same as (a), with the addition
of objects from the E2 and W2 fields.  (d) The same as (a), with the
addition of objects from the N2, S2, and SW2 fields.

Fig. 13. Gray-scale Hess diagrams for the CMDs in Figure~12.  Darker
shades correspond to a larger density of points. The contour in (a),
(c), and (d) is an isopleth from (b) that outlines the expected
principal sequences of Draco.  (a) The Hess diagram corresponding to
Figure~12(a).  (b) The same as (a) for the CMD in Figure~12(b). (c) The
same as (a) for the CMD in Figure~12(c). (d) The same as (a) for the
CMD in Figure~12(d).

Fig. 14. (a) Color-magnitude diagram for those quadrants of fields with
higher than average values of $\Sigma_{\rm in}$.  (b) Color-magnitude
diagram for those quadrants of fields with higher than average values
of $\Sigma_{\rm in}/\Sigma_{\rm out}$. (c) Gray-scale Hess diagram for
the CMD in (a). Darker shades correspond to a larger density of
points.  The contour is an isopleth from Figure~12(b) that outlines the
expected principal sequences of Draco.  (d) The same as (c) for
the CMD in (b).

Fig. 15. (a) The fractional rms scatter of $\Sigma_{\rm in}$ around its
mean as a function of limiting magnitude (solid squares) for the fields
N2, E2, S2, SW2, and W2, and those sections of the SE1 and W1 fields
outside of the measured tidal boundary of Draco.  $\Sigma_{\rm in}$
includes objects regardless of their CHI and SHARP values.  The open
squares represent the scatter expected from Poisson noise.  (b) The
same as (a) for $\Sigma_{\rm out}$.  (c) The same as (a) for
$\Sigma_{\rm in}/\Sigma_{\rm out}$.  (d) The same as (a) for
$\Sigma_{\rm in}$ including only objects with CHI values less than
5.0 and SHARP values within the limits shown in Fig.~8. (e) The same as
(d) for $\Sigma_{\rm out}$.  (f) The same as (d) for $\Sigma_{\rm
in}/\Sigma_{\rm out}$.

Fig. 16.  The values of $\Sigma_{\rm in}$, $\Sigma_{\rm out}$, and
$\Sigma_{\rm in}/\Sigma_{\rm out}$ to a limiting magnitude $R=22.6$
for the fields SE1, $W1^{*}$ (the section of W1 outside of the measured
tidal boundary of Draco), N2, E2, S2, SW2, and W2.  The top panels show
values calculated including objects regardless of their CHI and SHARP
values.  The error bars are $\sigma_{\rm int}$, the Poisson noise.  The
middle panels are the same as the top panels for values calculated including
objects with CHI values less than 5.0 and SHARP values within the
limits shown in Fig.~8.  The bottom panels are the same as the middle
panels, but with the error bars given $\sigma_{\rm ext}$ --- the rms
scatter around the mean value for the four quadrants of the field.  No
point is plotted for W1$^{*}$ since $\sigma_{\rm ext}$ cannot be
determined for this section of the W1 field.
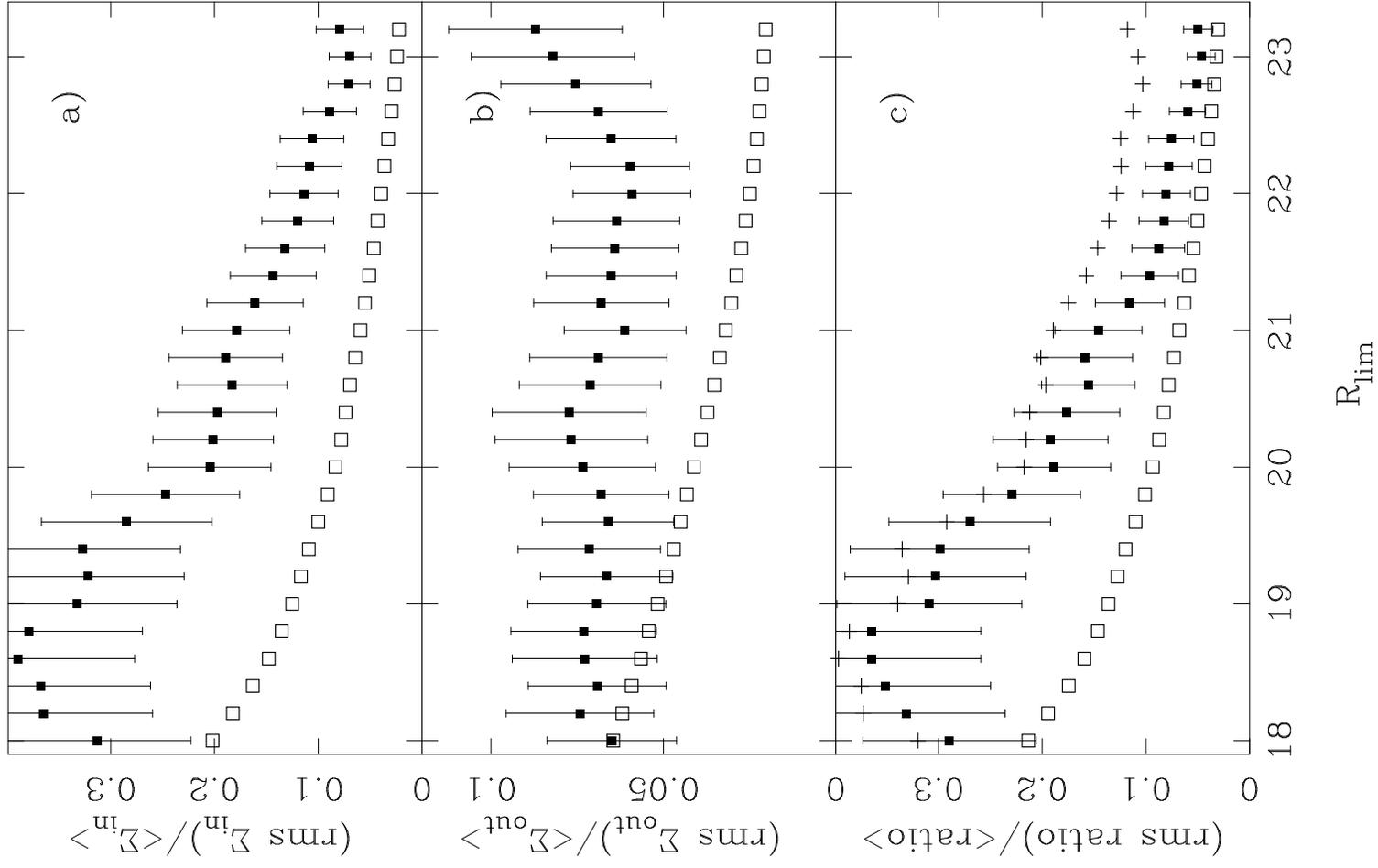
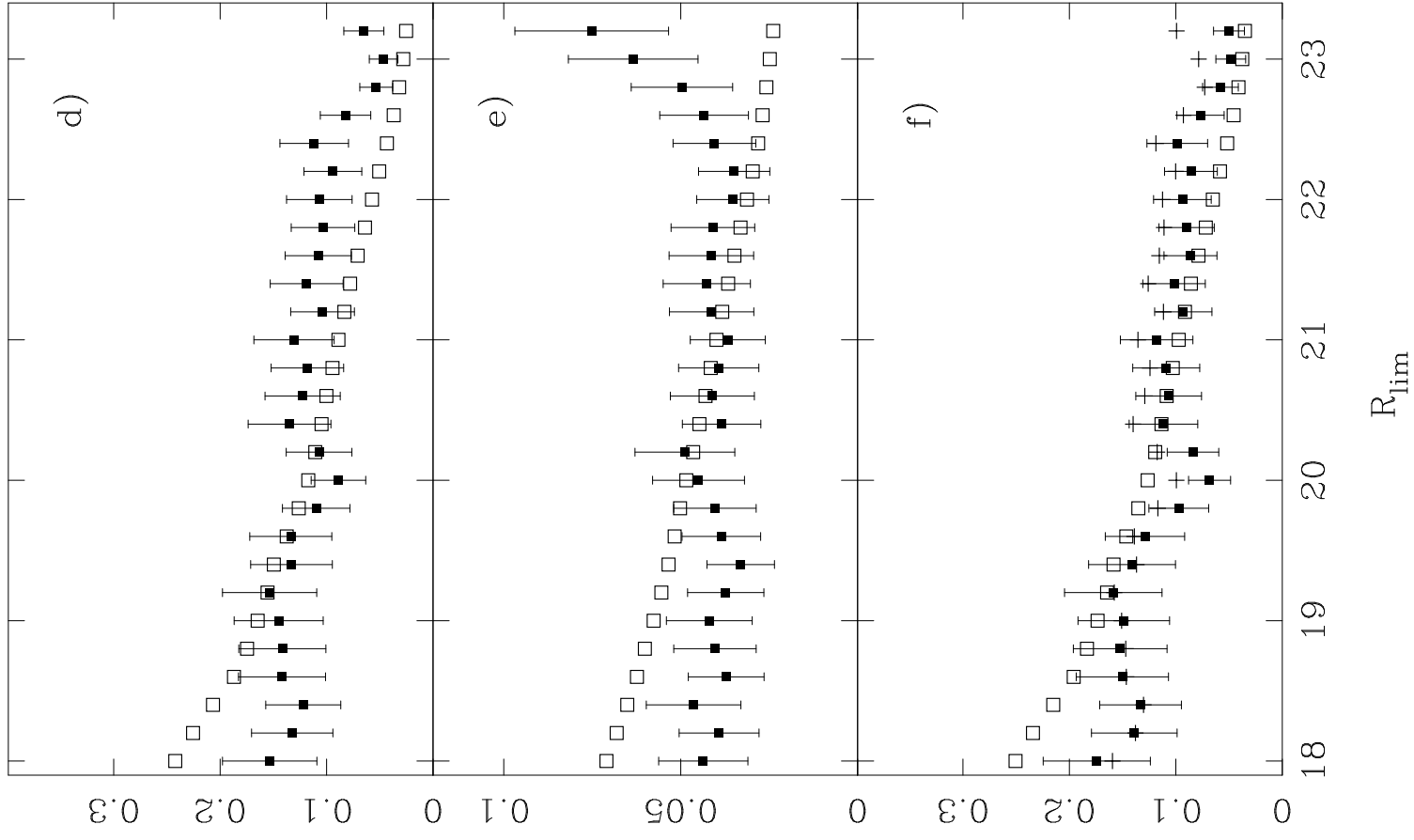
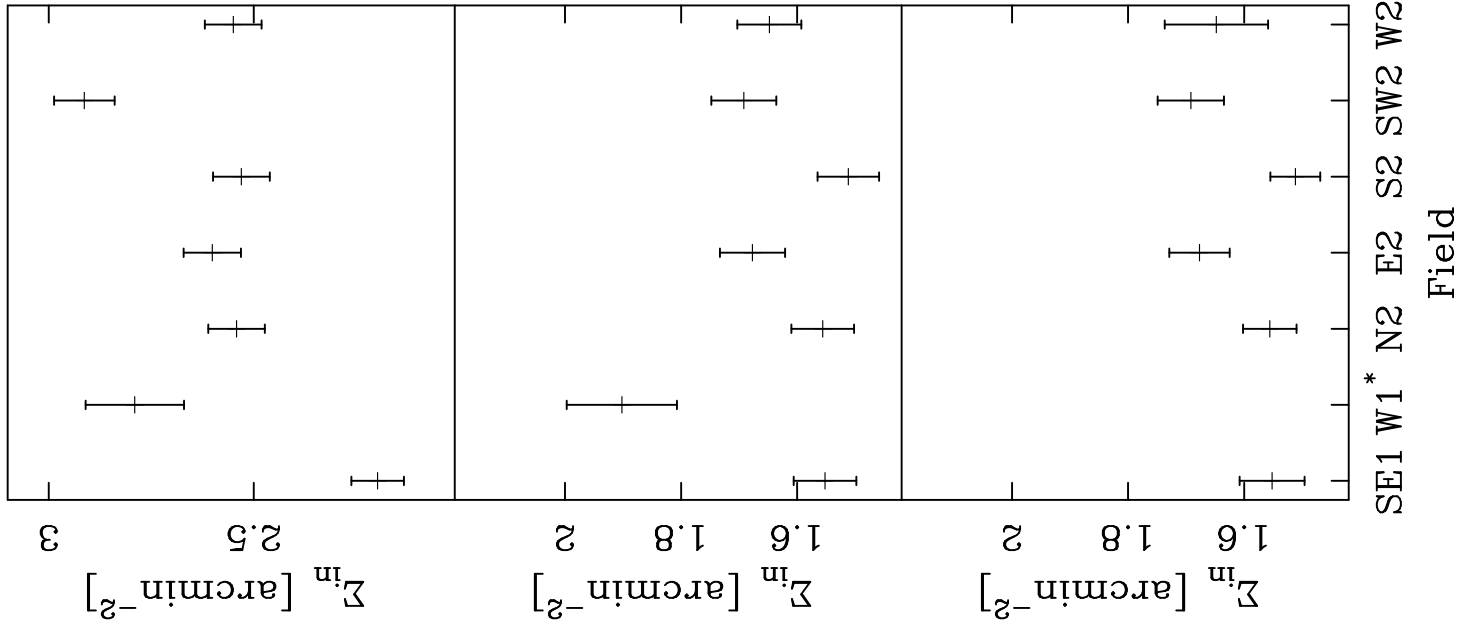
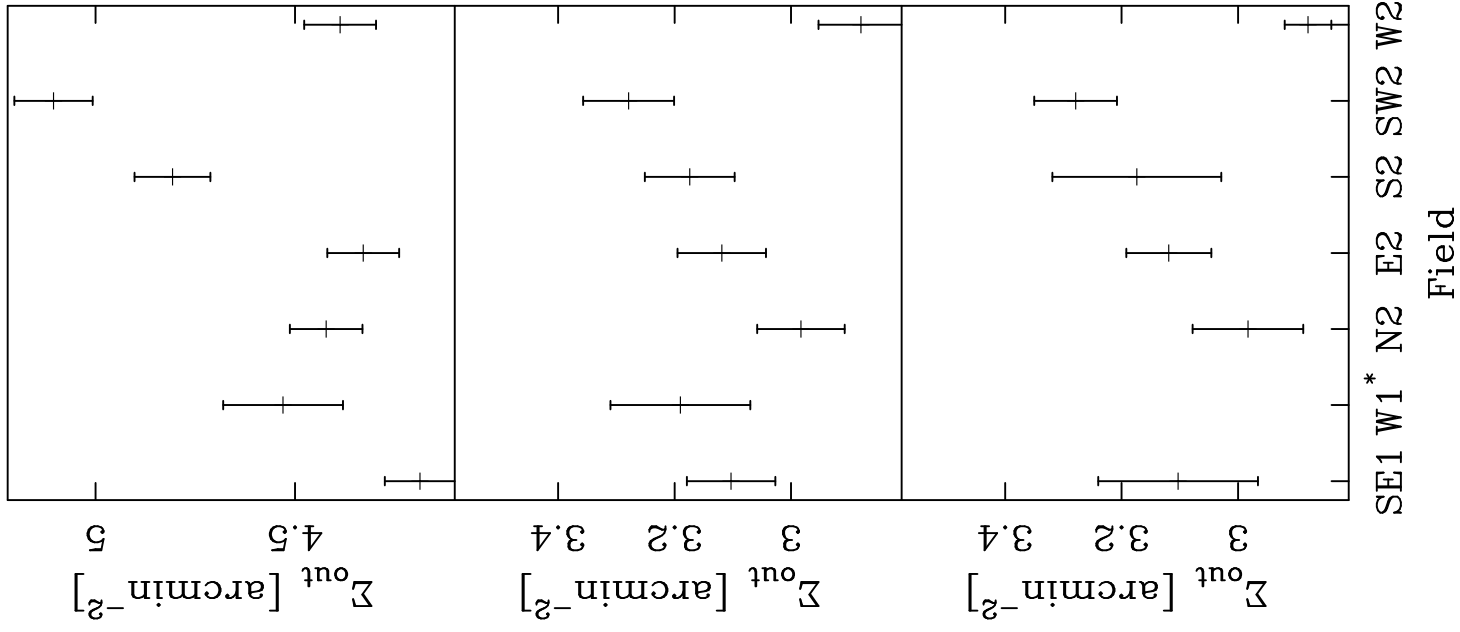
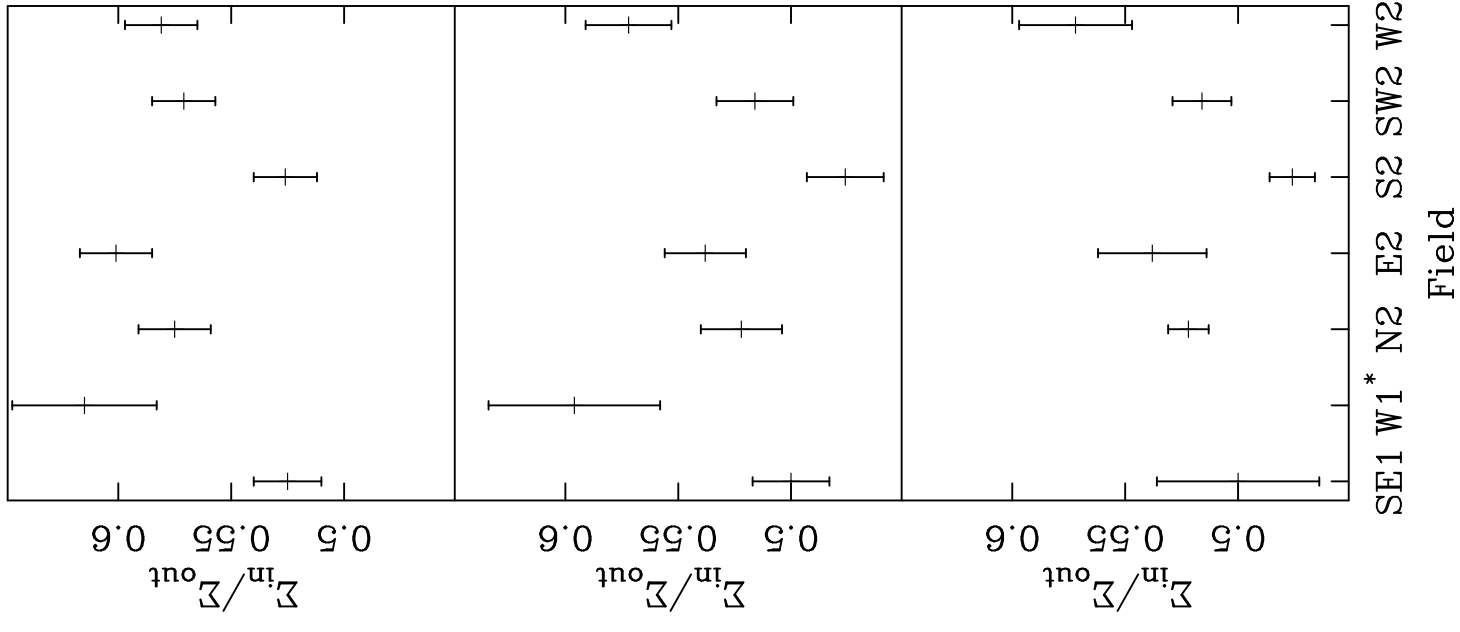
\end{document}